\documentclass[aps,prd,amsmath,showpacs,nofootinbib]{revtex4}
\RequirePackage[]{amssymb}
\usepackage{graphicx}

\def\({\left(}
\def\){\right)}

\newcommand{\beq}{\begin{equation}}
\newcommand{\eeq}{\end{equation}}
\newcommand{\bea}{\begin{eqnarray}}
\newcommand{\eea}{\end{eqnarray}}

\newcommand{\bean}{\begin{eqnarray*}}
\newcommand{\eean}{\end{eqnarray*}}
\newcommand{\bs}{\begin{subequations}}
\newcommand{\es}{\end{subequations}}

\newcommand{\Ref}[1]{(\ref{#1})}

\begin{document}


\title{Exact cosmological solutions with
nonminimal derivative coupling}

\author{Sergey V. Sushkov}%
\email{sergey_sushkov@mail.ru}%
\affiliation{Department of General Relativity and Gravitation, Kazan
State University, Kremlevskaya str. 18, Kazan 420008,
Russia} %
\affiliation{Department of Mathematics, Tatar State University of
Humanities and Education, Tatarstan str. 2, Kazan 420021, Russia}

\begin{abstract}
We consider a gravitational theory of a scalar field $\phi$ with
nonminimal derivative coupling to curvature. The coupling terms
have the form $\kappa_1 R\phi_{,\mu}\phi^{,\mu}$ and $\kappa_2
R_{\mu\nu}\phi^{,\mu}\phi^{,\nu}$ where $\kappa_1$ and $\kappa_2$
are coupling parameters with dimensions of length-squared. In
general, field equations of the theory contain third derivatives
of $g_{\mu\nu}$ and $\phi$. However, in the case
$-2\kappa_1=\kappa_2\equiv\kappa$ the derivative coupling term
reads $\kappa G_{\mu\nu}\phi^{,\mu}\phi^{,\nu}$ and the order of
corresponding field equations is reduced up to second one.
Assuming $-2\kappa_1=\kappa_2$, we study the spatially-flat
Friedman-Robertson-Walker model with a scale factor $a(t)$ and
find new exact cosmological solutions. It is shown that properties
of the model at early stages crucially depends on the sign of
$\kappa$. For negative $\kappa$ the model has an initial
cosmological singularity, i.e. $a(t)\sim (t-t_i)^{2/3}$ in the
limit $t\to t_i$; and for positive $\kappa$ the universe at early
stages has the quasi-de Sitter behavior, i.e. $a(t)\sim e^{Ht}$ in
the limit $t\to-\infty$, where $H=(3\sqrt{\kappa})^{-1}$. The
corresponding scalar field $\phi$ is exponentially growing at
$t\to-\infty$, i.e. $\phi(t)\sim e^{-t/\sqrt{\kappa}}$. At late
stages the universe evolution does not depend on $\kappa$ at all;
namely, for any $\kappa$ one has $a(t)\sim t^{1/3}$ at
$t\to\infty$. Summarizing, we conclude that a cosmological model
with nonminimal derivative coupling of the form $\kappa
G_{\mu\nu}\phi^{,\mu}\phi^{,\nu}$ is able to explain in a unique
manner both a quasi-de Sitter phase and an exit from it without
any fine-tuned potential.
\end{abstract}


 \maketitle

\section{Introduction\label{I}}
For many years scalar fields have been an object of great interest
for physicists. The reasons for this are manifold. One of them is
quite pragmatic: models with scalar fields are relatively simple,
and therefore it appeared possible to study them in detail and
then extrapolate the results to more realistic and complicated
models. More physical motivations include such important topics as
the idea about variable ``fundamental'' constants, the
Jordan-Brans-Dicke theory initially elaborated to solve the Mach
problem, the Kaluza-Klein compactification scheme, the low-energy
limit of the superstring theory, and others. Scalar fields play an
especially important role in cosmology. As a bright example, one
may mention numerous inflationary models in which inflation in the
early Universe is typically driven by a fundamental scalar field
called an inflaton. Furthermore, a recent discovery of cosmic
acceleration has only refreshed the interest to scalar fields
which began to be considered as candidates to explain dark energy
phenomena.

The rather general form of action for a scalar-tensor theory of
gravity with a single scalar field $\phi$ can be given
as\footnote{Throughout this paper we use units such that $G=c=1$.
The metric signature is $(- + + +)$ and the conventions for
curvature tensors are $R^\alpha_{\beta\gamma\delta} =
\Gamma^\alpha_{\beta\delta,\gamma} - ...$ and $ R_{\mu\nu} =
R^\alpha_{\mu\alpha\nu}$.}
\beq\label{STTaction}
S=\int d^4x\sqrt{-g}\left\{\frac1{8\pi}F(\phi,R)
-h(\phi)g_{\mu\nu}\phi^{,\mu}\phi^{,\nu}\right\},
\eeq
where $g_{\mu\nu}$ is a metric, $g=\det(g_{\mu\nu})$, and $R$ is
the scalar curvature. Functions $F(\phi,R)$ and $h(\phi)$ are
varying from theory to theory. The function $h(\phi)$ is
responsible for the sign of kinetic energy of the scalar field.
For example, the choice $h(\phi)\equiv-1$ leads to a wide class of
theories with the negative kinetic term. The function $F(\phi,R)$,
being in general nonlinear, provides a nonminimal coupling between
a scalar field and curvature. Though a freedom in choosing of
$F(\phi,R)$ leads to an unlimited variety of scalar-tensor
theories, it is known (see, for example,
\cite{Mae,FarGunNar,CroFra}) that there exist conformal
transformations transforming this kind of theories to Einstein's
theory with a new minimally coupled scalar field $\phi$ and an
effective potential $V(\phi)$ describing its self-interaction. The
potential $V(\phi)$ is a very important ingredient of various
cosmological models: a slowly varying potential behaves like an
effective cosmological constat providing one or more than one
inflationary phases. An appropriate choice of $V(\phi)$ is known
as a problem of fine tuning of the cosmological constant.

A further extension of scalar-tensor theories can be represented
by models with nonminimal couplings between derivatives of a
scalar field and curvature. This kind of couplings may appear in
some Kaluza-Klein theories \cite{KK1,KK2} (see also
\cite{Lindebook}, Section 9.5). In 1993, Amendola \cite{Ame} has
been considered the most general gravity Lagrangian linear in the
curvature scalar $R$, quadratic in $\phi$, and containing terms
with four derivatives including all of the following terms (see
also \cite{CapLamSch} for details):
\bea
&
L_1=\kappa_1 R\phi_{,\mu}\phi^{,\mu};\quad%
L_2=\kappa_2 R_{\mu\nu}\phi^{,\mu}\phi^{,\nu};\quad%
L_3=\kappa_3 R \phi\square\phi ; &\nonumber\\
&
L_4=\kappa_4 R_{\mu\nu} \phi\phi^{;\mu\nu} ;\quad%
L_5=\kappa_5 R_{;\mu} \phi\phi^{,\mu} ;\quad%
L_6=\kappa_6 \square R \phi^2, &\nonumber
\eea
where coefficients
$\kappa_1,\dots,\kappa_6$ are coupling parameters with dimensions
of length-squared. Using the divergencies
$$
(R\phi^{,\mu}\phi)_{;\mu};\quad
(R^{\mu\nu}\phi\phi_{,\mu})_{;\nu};\quad%
(R^{;\mu}\phi^2)_{;\mu},
$$
one may conclude that, without loss of generality, $L_4$, $L_5$,
and $L_6$ are not necessary to be considered. Also one may rule
out $L_3$ because it contains $\phi$ itself, while coupling term
of main interest are those, where only the gradient of $\phi$ is
included. Thus, a general scalar-tensor theory with nonminimal
derivative couplings may include only two terms $L_1$ and $L_2$.

As was shown by Amendola \cite{Ame}, a theory with derivative
couplings cannot be recasting into Einsteinian form by a conformal
rescaling $\tilde g_{\mu\nu}=e^{2\omega}g_{\mu\nu}$. He also
supposed that an effective cosmological constant, and then the
inflationary phase can be recovered without considering any
effective potential if a nonminimal derivative coupling is
introduced. Amendola himself \cite{Ame} has considered a
cosmological model in the theory with the only derivative coupling
term $L_1=\kappa_1 R\phi_{,\mu}\phi^{,\mu}$ and, by using a
generalized slow-rolling approximation (i.e., neglecting all terms
of order higher than the second one), he has obtained some
analytical inflationary solutions. A general model containing both
$L_1$ and $L_2$ has been discussed in \cite{CapLamSch} (see also
\cite{CapLam}); it was shown that the de Sitter spacetime is an
attractor solution of the model if $4\kappa_1+\kappa_2>0$.
Recently Daniel and Caldwell \cite{DanCal} have considered a
theory with the derivative coupling term $L_2=\kappa_2
R_{\mu\nu}\phi^{,\mu}\phi^{,\nu}$; in particular, they studied
constraints which precision tests of general relativity impose on
the coupling parameter $\kappa_2$. It is also worth mentioning a
series of papers devoted to a nonminimal modification of the
Einstein-Yang-Mills-Higgs theory \cite{BalDehZay:07} (see also a
review \cite{BalDehZay} and references therein).

In this paper we continue studying a scalar-tensor theory with
nonminimal derivative couplings and construct new exact
cosmological solutions of the theory.

\section{Field equations\label{II}}
Let us consider a gravitational theory of a scalar field $\phi$
with nonminimal derivative couplings to curvature described by the
action
\beq\label{action}
S=\int d^4x\sqrt{-g}\left\{\frac{R}{8\pi}
-\big[g_{\mu\nu}+\kappa_1 g_{\mu\nu} R+\kappa_2
R_{\mu\nu}\big]\phi^{,\mu}\phi^{,\nu} \right\}.
\eeq
Here coefficients $\kappa_1$ and $\kappa_2$ are derivative
coupling parameters with dimensions of length-squared. Note that
the action \Ref{action} does not include any potential. Varying
the action \Ref{action} with respect to the metric $g_{\mu\nu}$
gives the gravitational field equations
\beq\label{eineqgen}
G_{\mu\nu}=8\pi\big[T_{\mu\nu} +\kappa_1
\Theta^{(1)}_{\mu\nu}+\kappa_2\Theta^{(2)}_{\mu\nu}\big],
\eeq
with
\bea
T_{\mu\nu}&=&\nabla_\mu\phi\nabla_\nu\phi-
{\textstyle\frac12}g_{\mu\nu}(\nabla\phi)^2-g_{\mu\nu} V(\phi),
\nonumber\\
\Theta^{(1)}_{\mu\nu}&=& \nabla_\mu\phi\nabla_\nu\phi R
+(\nabla\phi)^2 G_{\mu\nu} -\nabla_\mu\nabla_\nu(\nabla\phi)^2
+g_{\mu\nu}\square(\nabla\phi)^2,
\nonumber\\
\Theta^{(2)}_{\mu\nu}&=&-{\textstyle\frac12}
g_{\mu\nu}\nabla_\alpha\phi \nabla_\beta\phi R^{\alpha\beta} +
2\nabla_\alpha\phi \nabla_{(\mu}\phi
R_{\nu)}^\alpha+{\textstyle\frac12}\square(\nabla_\mu\phi\nabla_\nu\phi)
-\nabla_\alpha\nabla_{(\mu}(\nabla_{\nu)}\phi\nabla^\alpha\phi)
\nonumber\\
&&+{\textstyle\frac12}
g_{\mu\nu}\nabla_\alpha\nabla_\beta(\nabla^\alpha\phi\nabla^\beta\phi).
\nonumber
\eea
where $G_{\mu\nu}=R_{\mu\nu}-\frac12 g_{\mu\nu}R$ is the Einstein
tensor. Then, varying the action \Ref{action} with respect to
$\phi$ gives the scalar field equation of motion:
\beq\label{eqmogen}
g^{\mu\nu}\nabla_\mu\nabla_\nu\phi+
\nabla_{\mu}\big[\nabla_\nu\phi(\kappa_1 g^{\mu\nu} R+\kappa_2
R^{\mu\nu})\big]=0.
\eeq
Note also that because of the Bianchi identity, $\nabla_\mu
G^{\mu\nu}=0$, the scalar field and order-$\kappa$ terms form a
conserved system, hence the scalar field equation \Ref{eqmogen}
can be obtained as a consequence of the generalized conservation
law $\nabla_\mu[T^{\mu\nu}
+\kappa_1\Theta^{(1)\mu\nu}+\kappa_2\Theta^{(2)\mu\nu}]=0$.

Generally, the gravitational field equations \Ref{eineqgen}
contain third derivatives of $\phi$, whilst the scalar field
equation \Ref{eqmogen} contains third derivatives of the metric.
However, an important feature of the theory \Ref{action} is the
fact that the order of field equations can be reduced for a
specific choice of $\kappa_1$ and $\kappa_2$. To show this we
rewrite, after some algebra, the expressions for
$\Theta^{(1)}_{\mu\nu}$ and $\Theta^{(2)}_{\mu\nu}$ as follows:
\bea
\Theta^{(1)}_{\mu\nu}&=&\nabla_\mu\phi\,\nabla_\nu\phi\,R
+(\nabla\phi)^2 G_{\mu\nu}
-2\nabla^\alpha\phi\,\nabla^\beta\phi\,R_{\mu\alpha\nu\beta}-2\nabla_\mu\nabla^\alpha\phi\,\nabla_\nu\nabla_\alpha\phi
-2\nabla^\alpha\phi\,\nabla_\alpha\nabla_\mu\nabla_\nu\phi
\nonumber\\
&&+2g_{\mu\nu}\big[\nabla^\alpha\nabla^\beta\phi\,\nabla_\alpha\nabla_\beta\phi
+\nabla_\alpha\phi\,\nabla_\beta\phi\,R^{\alpha\beta}
+\nabla^\alpha\phi\,\nabla_\alpha\square\phi \big],\\
\Theta^{(2)}_{\mu\nu}&=&2\nabla_\alpha\phi\,\nabla_{(\mu}\phi
R^\alpha_{\nu)} -\nabla_\mu\nabla_\nu\phi\,\square\phi
-\nabla^\alpha\phi\,\nabla_\alpha\nabla_\mu\nabla_\nu\phi
\nonumber\\
&&
+{\textstyle\frac12}g_{\mu\nu}\big[\nabla^\alpha\nabla^\beta\phi\,\nabla_\alpha\nabla_\beta\phi
+(\square\phi)^2 +2\nabla^\alpha\phi\,\nabla_\alpha\square\phi
\big].
\eea
It is seen that both expressions contain similar third-order
terms, $\nabla^\alpha\phi\,\nabla_\alpha\nabla_\mu\nabla_\nu\phi$
and $g_{\mu\nu}\nabla^\alpha\phi\,\nabla_\alpha\square\phi$, which
are cancelled in the combination $\kappa_1
\Theta^{(1)}_{\mu\nu}+\kappa_2\Theta^{(2)}_{\mu\nu}$ provided the
coupling parameters are chosen as follows
\beq\label{kappa}
-2\kappa_1=\kappa_2.
\eeq

Hereafter we will assume that the relation \Ref{kappa} hold.
Denoting $\kappa=-2\kappa_1=\kappa_2$, we can rewrite the action
\Ref{action} as follows
\beq\label{action2}
S=\int d^4x\sqrt{-g}\left\{\frac{R}{8\pi} - [g_{\mu\nu}+\kappa
G_{\mu\nu}]\phi^{,\mu}\phi^{,\nu}\right\}.
\eeq
The gravitational field equations \Ref{eineqgen} now read
\beq\label{eineq}
G_{\mu\nu}=8\pi\big[T_{\mu\nu} +\kappa \Theta_{\mu\nu}\big],
\eeq
with
\bea
\Theta_{\mu\nu}&=&-{\textstyle\frac12}\Theta^{(1)}_{\mu\nu}+\Theta^{(2)}_{\mu\nu}
\nonumber\\&=&-{\textstyle\frac12}\nabla_\mu\phi\,\nabla_\nu\phi\,R
+2\nabla_\alpha\phi\,\nabla_{(\mu}\phi R^\alpha_{\nu)}
-{\textstyle\frac12}(\nabla\phi)^2 G_{\mu\nu}
+\nabla^\alpha\phi\,\nabla^\beta\phi\,R_{\mu\alpha\nu\beta}+\nabla_\mu\nabla^\alpha\phi\,\nabla_\nu\nabla_\alpha\phi
\nonumber\\
&& -\nabla_\mu\nabla_\nu\phi\,\square\phi
+g_{\mu\nu}\big[-{\textstyle\frac12}\nabla^\alpha\nabla^\beta\phi\,\nabla_\alpha\nabla_\beta\phi
+{\textstyle\frac12}(\square\phi)^2
-\nabla_\alpha\phi\,\nabla_\beta\phi\,R^{\alpha\beta} \big],
\nonumber
\eea
and the scalar field equation of motion \Ref{eqmogen} yields
\beq\label{eqmo}
[g^{\mu\nu}+\kappa G^{\mu\nu}]\nabla_{\mu}\nabla_\nu\phi=0.
\eeq

Let us emphasize once more that the field equations \Ref{eineq}
and \Ref{eqmo} contain now only second derivatives of $g_{\mu\nu}$
and $\phi$. Thus, from the physical point of view, the theory
\Ref{action2} can be interpreted as a ``good'' dynamical theory.

\section{Cosmological models}

Consider a spatially-flat cosmological model with a metric
\beq\label{metric}
ds^2=-dt^2+e^{2\alpha(t)}d{\rm\bf x}^2,
\eeq
where $a(t)=e^{\alpha(t)}$ is the scale factor, and $d{\rm\bf
x}^2$ is the Euclidian metric, and assume that $\phi=\phi(t)$. In
this case the field equations \Ref{eineq} and \Ref{eqmo} are
reduced to the following system:
\bea\label{00compt}
&3\dot{\alpha}^2=4\pi\dot{\phi}^2\left(1-9\kappa\dot{\alpha}^2\right),&
\\ \label{11compt}
&\displaystyle -2\ddot{\alpha}-3\dot{\alpha}^2=4\pi\dot{\phi}^2
\left[1+\kappa\left(2\ddot{\alpha}+3\dot{\alpha}^2+4\dot{\alpha}\ddot{\phi}\dot{\phi}^{-1}\right)\right],&
\\ \label{eqmocosmo}
&\ddot\phi+3\dot\alpha\dot\phi-3\kappa\left[\dot\alpha^2\ddot\phi
+2\dot\alpha\ddot\alpha\dot\phi+3\dot\alpha^3\dot\phi\right]=0.&
\eea
where a dot means a derivative with respect to time. Note that
Eqs.~\Ref{11compt} and \Ref{eqmocosmo} are of second order, while
Eq.~\Ref{00compt} is a first-order differential constraint for
$\alpha(t)$ and $\phi(t)$.

First, let us discuss the simple case $\kappa=0$, which just means
the absence of derivative coupling. In this case Eqs.
\Ref{00compt}-\Ref{eqmocosmo} are easily solved resulting in
\bea\label{alpha0}
\alpha(t)&=&\alpha_0+\frac13\ln(t-t_0),
\\\label{phi0}
\phi(t)&=&\phi_0+\frac1{2\sqrt{3\pi}}\ln(t-t_0),
\eea
where $t_0$, $\alpha_0$ and $\phi_0$ are constants of integration.
Without loss of generality one may put $\alpha_0=0$ and
$\phi_0=0$, then the corresponding metric reads
\beq\label{metric0}
ds^2=-dt^2+(t-t_0)^{2/3}d{\rm\bf x}^2.
\eeq
The spacetime with the metric \Ref{metric0} has an initial
singularity at $t=t_0$.

Consider now a general case $\kappa\not=0$. In this case the
constraint \Ref{00compt} can be rewritten as follows
\beq\label{constrphi}
\dot\phi^2=\frac{3\dot\alpha^2}{4\pi(1-9\kappa\dot\alpha^2)},
\eeq
or, equivalently,
\beq\label{constralpha}
\dot\alpha^2=\frac{4\pi\dot\phi^2}{3(1+12\pi\kappa\dot\phi^2)}.
\eeq
From here it follows that $\dot\alpha$ and $\dot\phi$ should obey
the following conditions:
\bea
&1-9\kappa\dot\alpha^2>0,&
\label{ineqalpha}\\
&1+12\pi\kappa\dot\phi^2>0.& \label{ineqphi}
\eea
Let us now separate equations for $\alpha$ and $\phi$. For this
aim, we resolve Eqs. \Ref{11compt} and \Ref{eqmocosmo} with
respect to $\ddot\alpha$ and $\ddot\phi$ and, using the relations
\Ref{constrphi} and \Ref{constralpha}, eliminate $\dot\phi$ and
$\dot\alpha$ from respective equations. As the result, we find
\beq\label{a2}
    \ddot\alpha=-\frac{3\dot\alpha^2(1-3\kappa\dot\alpha^2)(1-9\kappa\dot\alpha^2)}
    {1-9\kappa\dot\alpha^2+54\kappa^2\dot\alpha^4},
\eeq
\beq\label{phi2}
    \ddot\phi=-\frac{2\sqrt{3\pi}\dot\phi^2(1+
    8\pi\kappa\dot\phi^2)(1+12\pi\kappa\dot\phi^2)^{1/2}}
    {1+12\pi\kappa\dot\phi^2+96\pi^2\kappa^2\dot\phi^4}.
\eeq
Since $\dot\alpha$ and $\dot\phi$ obey the conditions
\Ref{ineqalpha} and \Ref{ineqphi}, it is seen that $\ddot\alpha$
and $\ddot\phi$ are {negative} for {\em all} times. In turns, this
means that $\dot\alpha$ and $\dot\phi$ are monotonically
decreasing with time.

Let us analyze an asymptotical behavior of $\alpha$ and $\phi$ for
large times. Suppose that $\dot\alpha$ tends to some nonzero
constant at $t\to\infty$; respectively, this means that
$\ddot\alpha$ should go to zero. However, as it follows from Eq.
\Ref{a2}, $\ddot\alpha$ is not zero in this limit. Thus, we face
with a contradiction and, therefore, should conclude that
$\dot\alpha\to 0$ if $t\to \infty$. By using this asymptotical
property, we obtain the following asymptotic form of Eq. \Ref{a2}:
\beq
\ddot\alpha \approx -3\dot\alpha^2.
\eeq
The corresponding asymptotical solution is
\beq\label{alphaasymp}
\alpha_{t\to\infty} = \alpha_{\infty}+\frac13\ln(t-t_\infty),
\eeq
where $t_\infty$ and $\alpha_\infty$ are constants of integration.
An asymptotic for $\phi$ can be found straightforwardly from the
constraint \Ref{constrphi}:
\beq\label{phiasymp}
\phi_{t\to\infty}=\phi_\infty+\frac1{2\sqrt{3\pi}}\ln(t-t_\infty),
\eeq
where $\phi_\infty$ is a constant of integration. It is worth
noting that the asymptotics \Ref{alphaasymp} and \Ref{phiasymp} do
{\em not} depend on $\kappa$ and coincide with exact solutions
\Ref{alpha0} and \Ref{phi0} obtained for $\kappa=0$.

To characterize an asymptotical behavior of $\alpha$ and $\phi$
for small times, we consider separately two cases.

First, let $\kappa$ be negative, $\kappa<0$. In this case the
condition \Ref{ineqphi} gives the following bound for $\dot\phi$:
\beq\label{ineqphi1}
\dot\phi^2<\frac1{12\pi|\kappa|},
\eeq
while the condition \Ref{ineqalpha} is fulfilled for all values of
$\dot\alpha$. Since $\dot\alpha$ is monotonically decreasing with
time, its value should be growing with decreasing time. Let $t_i$
be some initial moment of time (possibly $t_i=-\infty$). Suppose
that $\dot\alpha$ tends to some constant value in the limit $t\to
t_i$; respectively, this means that $\ddot\alpha$ should go to
zero. However, as follows from Eq. \Ref{a2}, $\ddot\alpha$ is not
zero in this limit. This is a contradiction, and hence we should
conclude that $\dot\alpha$ is boundlessly increasing in the limit
$t\to t_i$. Assuming $\dot\alpha\to\infty$ at $t\to t_i$ gives the
following asymptotical form of Eq. \Ref{a2}:
\beq
\ddot\alpha \approx -\frac32\dot\alpha^2,
\eeq
with the asymptotic solution
\beq\label{initasalpha}
\alpha_{t\to t_i} = \alpha_{i}+\frac23\ln(t-t_i).
\eeq
The asymptotic for $\phi$ is found from Eq. \Ref{constrphi} as
\beq\label{phi_k<0}
\phi_{t\to t_i}=\phi_i+\frac{t}{2\sqrt{3\pi|\kappa|}},
\eeq
where $t_i$, $\alpha_i$ and $\phi_i$ are constants of integration.
The corresponding asymptotical form of the metric \Ref{metric} is
\beq\label{metric_k<0}
ds^2_{t\to t_i} =-dt^2+e^{2\alpha_i}(t-t_i)^{4/3} d{\rm\bf x}^2.
\eeq
A spacetime with this metric is singular at $t=t_i$. This
singularity is analogous to initial cosmological singularities in
models with usual scalar fields. However, a new interesting
feature of the examined model is that the scalar field with
negative derivative coupling $\kappa<0$ has the regular behavior
\Ref{phi_k<0} near the singularity.

\begin{figure}[t]
\centerline{\hbox{\includegraphics[width=7cm]{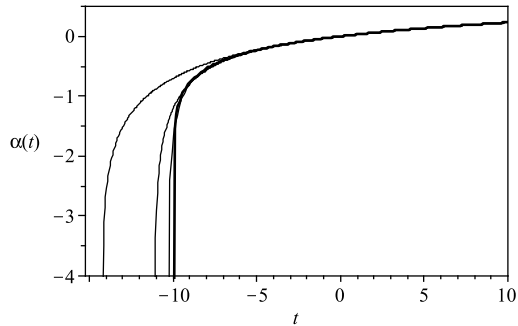}} }
\caption{\label{f1} Plots of $\alpha(t)$ for negative
$\kappa=-1;-10;-100$ (from right to left). The thick curve
corresponds to $\kappa=0$ (no derivative coupling).}
\end{figure}

Results of numerical study of Eq. \Ref{a2} in case $\kappa<0$ are
shown in Fig. \ref{f1}. Obtained solutions reproduce all
asymptotical properties found above analytically.

Then, let $\kappa$ be positive, $\kappa>0$. In this case the
condition \Ref{ineqalpha} gives the following bound for
$\dot\alpha$:
\beq\label{ineqalpha1}
\dot\alpha^2<\frac1{9\kappa},
\eeq
while Eq. \Ref{ineqphi} is fulfilled for any $\dot\phi$. Repeating
the above arguments, we may conclude that $\dot\phi\to\infty$ at
$t\to-\infty$. This gives the following asymptotic form of Eq.
\Ref{phi2}:
\beq
\ddot\phi\approx-\frac{\dot\phi}{\sqrt{\kappa}}
\eeq
with the asymptotic solution
\beq\label{asympphi}
\phi_{t\to-\infty} = \phi_2+C e^{-t/\sqrt{\kappa}},
\eeq
where $\phi_2$ and $C$ are constants of integration.
To obtain an asymptotic for $\alpha$ one may substitute Eq.
\Ref{asympphi} into \Ref{constralpha} and find after some algebra
\beq\label{a_k>0}
\alpha\approx
\frac{t-t_0}{3\sqrt{\kappa}}-\frac{e^{2t/\sqrt{\kappa}}}{144\pi\kappa
C^2},
\eeq
where $t_0$ is a constant of integration which without loss of
generality can be set zero. We see that in the limit $t\to-\infty$
the scalar $\phi$ is exponentially growing, and $\alpha$ is
exponentially approximating to its asymptotic
\beq
\alpha_{t\to-\infty}=\frac{t}{3\sqrt{\kappa}}.
\eeq
\begin{figure}[t]\label{fig1}
\centerline{\hbox{\includegraphics[width=7cm]{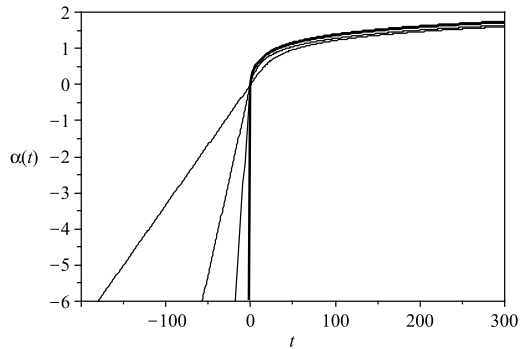}}}
\caption{\label{f2} Plots of $\alpha(t)$ for positive
$\kappa=1;10;100$ (from right to left). The thick curve
corresponds to $\kappa=0$ (no derivative coupling).}
\end{figure}
Hence, in the limit $t\to-\infty$ the spacetime metric
\Ref{metric} takes asymptotically the de Sitter-like form:
\beq\label{deSitter}
ds^2_{t\to-\infty}=-dt^2+e^{2Ht} d{\rm\bf x}^2,
\eeq
with $H=(3\sqrt{\kappa})^{-1}$. Thus, in the case $\kappa>0$ an
universe at early stages has the quasi-de Sitter behavior
corresponding to the cosmological constant
$\Lambda=3H^2=(3\kappa)^{-1}$.

Results of numerical study in the case $\kappa>0$ are shown in
Fig. \ref{f2}. Note that all $\kappa>0$ solutions represent an
interesting feature. Namely, they describe two phases in evolution
of the universe. First, for an infinitely long time the universe
is living in the quasi-de Sitter or inflationary phase. Then,
during a relatively short time the universe exits from the
inflationary stage and goes to a power-law expansion with
$a(t)\sim t^{1/3}$ (it is worth noting that this law corresponds
to the equation of
state $p=\rho$). 


\section{Conclusions}
We have considered the gravitational theory of a scalar field with
nonminimal derivative coupling to curvature and studied
cosmological models in this theory. The main results obtained are
as follows:

1. The Lagrangian of the theory includes two derivative coupling
terms $\kappa_1 R\phi_{,\mu}\phi^{,\mu}$ and $\kappa_2
R_{\mu\nu}\phi^{,\mu}\phi^{,\nu}$, where $\kappa_1$ and $\kappa_2$
are coupling parameters with dimensions of length-squared. In
general, field equations of the theory are of third order, i.e.,
contain third derivatives of $g_{\mu\nu}$ and $\phi$, but in the
particular case the order of equations is reduced up to the second
one. This case corresponds to the choice
$-2\kappa_1=\kappa_2\equiv\kappa$, then a combination of
derivative coupling terms turn into $\kappa
G_{\mu\nu}\phi^{,\mu}\phi^{,\nu}$. It is worth noting that
Capozziello et al \cite{CapLamSch} , at pages 43 and 47, have
mentioned the case $-2\kappa_1=\kappa_2$ to play a special role,
because it represents a singular point of the differential
equation.
In this paper, we have supposed that the theory with
$-2\kappa_1=\kappa_2$ is more preferable with the physical point
of view, since the corresponding field equations do not contain
derivatives of dynamical variables of order higher than the
second.

2. Assuming $-2\kappa_1=\kappa_2\equiv\kappa$, we have studied a
cosmological model with the spatially-flat
Friedman-Robertson-Walker metric. It was shown that a behavior of
the scale factor $a(t)$ and the scalar field $\phi$ at large times
is the same for all values of $\kappa$ including zero, that is the
late evolution of universe does not depend on $\kappa$. Namely,
one has $a(t)\sim t^{1/3}$ and $\phi(t)\sim\ln t$ at $t\to\infty$.
Note this asymptotical behavior coincides with that of the exact
solution \Ref{alpha0}, \Ref{phi0} obtained for $\kappa=0$ (no
coupling).

3. General properties of the model crucially depends on a sign of
$\kappa$. For $\kappa<0$ an asymptotical form of the cosmological
metric for small times is given by Eq. \Ref{metric_k<0}. A
corresponding scale factor is $a(t)\sim (t-t_i)^{2/3}$; it
describes the universe with an initial singularity at $t=t_i$. A
new interesting feature of the model with derivative coupling is
that a behavior of the scalar field near the cosmological
singularity is regular, $\phi(t)\sim t$ (see Eq. \Ref{phi_k<0}).
For $\kappa>0$ the law of universe evolution is qualitatively
distinct from that for $\kappa<0$. Now at early stages the
universe has the quasi-de Sitter behavior \Ref{deSitter}
corresponding to the cosmological constant
$\Lambda=(3\kappa)^{-1}$. In the limit $t\to-\infty$ the scale
factor has the following asymptotical form $a(t)\sim
\exp\left(\frac{t}{3\kappa^{1/2}}-\frac{e^{2t/\sqrt{\kappa}}}{144\pi\kappa
C^2}\right)$ (see Eq. \Ref{a_k>0}), hence $a(t)$ exponentially
fast goes to the de-Sitter form $a(t)=e^{Ht}$ with
$H=(3\sqrt{\kappa})^{-1}$. At the same time, the scalar field
$\phi$ is exponentially growing at $t\to-\infty$, namely
$\phi(t)\sim e^{-t/\sqrt{\kappa}}$.

In conclusion, let us summarize the most essential features of
cosmology with nonminimal derivative coupling. First of all, we
should emphasize that cosmological solutions with the quasi-de
Sitter phase are typical solutions of the gravitational theory of
a scalar field with derivative coupling of the form $\kappa
G_{\mu\nu}\phi^{,\mu}\phi^{,\nu}$ with positive $\kappa$. So, in
order to obtain an inflationary phase, one need no fine-tuned
potential, and so one do not face with the problem of fine-tuning.
Another important feature of the model consists in the fact that
an exact cosmological solution with $\kappa>0$ describes in a
unique manner both a quasi-de Sitter phase and an exit from it.
Thus, the problem of graceful exit from inflation in cosmology
with the derivative coupling term $\kappa
G_{\mu\nu}\phi^{,\mu}\phi^{,\nu}$ has a natural solution without
any fine-tuned potential.


\subsection*{Acknowledgments}
This work was supported in part by the Russian Foundation for
Basic Research grants No. 08-02-91307, 08-02-00325.

\end{document}